\documentclass[aps,twocolumn,groupedaddress,reprint,amsmath,amssymb]{revtex4-2}
\usepackage[utf8]{inputenc}
\usepackage[T1]{fontenc}
\usepackage{amsmath}
\usepackage{amsfonts}
\usepackage{amssymb}
\usepackage{mathrsfs}
\usepackage{graphics,graphicx}
\usepackage{xcolor}
\usepackage{bm}
\usepackage[breaklinks=true]{hyperref}
\usepackage{setspace} 
\usepackage{caption}    
\usepackage{booktabs}   
\usepackage{multirow}
\usepackage{subcaption}
\usepackage{array}
\usepackage{titlesec}

\titlespacing*{\section}{0pt}{20pt}{10pt}  
\newcolumntype{M}[1]{>{\centering\arraybackslash}m{#1}} 

\begin{document}
	
		\title{Mass inflation and strong cosmic censorship conjecture in {the covariant quantum black hole}}
		
        \author{Jianhui Lin\textsuperscript{1}} 
        \author{Xiangdong Zhang\textsuperscript{1}}\email{Corresponding author. scxdzhang@scut.edu.cn}
        \author{Mois\'es Bravo-Gaete\textsuperscript{2}}
        \address{\textsuperscript{1}School of Physics and Optoelectronics, South China University of Technology, Guangzhou 510641, China, 
	\\\textsuperscript{2}Departamento de Matem\'atica, F\'isica y Estad\'istica, Facultad de Ciencias B\'asicas, Universidad Cat\'olica del Maule, Casilla 617, Talca, Chile.}

	\begin{abstract}	
			Recently, two types of solutions to the long-standing issue of general covariance in canonical quantum gravity have been proposed. From the above, a fundamental question arises: which solution is superior? Considering one type of solution with a Cauchy horizon, in the present letter, we explore whether it exhibits properties similar to those of the Reissner-Nordstr\"{o}m black hole. Due to its geometric similarity, application of the generalized Dray-'t Hooft-Redmond relation reveals evidence of mass inflation, indicating that the Cauchy horizon is unstable. While this is consistent with the Strong Cosmic Censorship conjecture, it suggests that it does not represent a regular black hole. Furthermore, we extend the metric to include a cosmological constant and study the validity of the Strong Cosmic Censorship conjecture for the quantum black hole in de Sitter spacetime. {After taking into account the reasonable ranges of parameters for the quantum black hole, we find that }the Strong Cosmic Censorship Conjecture holds.
	\end{abstract}
    \maketitle

\section{Introduction} \label{introduction}
Loop Quantum Gravity (LQG), as one of the candidates to avoid spacetime singularities \cite{Penrose1965}, is characterized by its background independence and non-perturbative formulation (see, e.g., \cite{Ashtekar2004, Thiemann2007, Han2005}). Over the past decade, numerous loop quantum symmetry-reduced models have been proposed \cite{Gambini2008, Ashtekar2018, Zhang2020, Kelly2020, Zhang2022,Lin2024,Shi2024,Zhang2023review}, within which the covariance problem is implicitly present in these effective theories. This problem is common in effective Hamiltonian theories resulting from canonical quantum gravity, meaning that physical predictions may not necessarily be independent of the choice of coordinates. Recently, conditions for the Hamiltonian constraint that ensure general covariance were provided in \cite{Zhang2024, Bojowald2024,Zhang2024-12}. Two distinct black hole (BH) solutions have been obtained through different polymerization schemes: one resembles the Reissner-Nordstr\"{o}m (RN) BH and features a Cauchy horizon \cite{Zhang2024}, while the other does not \cite{Zhang2024-12}. This topic has garnered increasing attention, with many related issues being explored \cite{Konoplya2024, Malik2024, Wang2025,Heidari2025, Zhu2025, Lan2024, Lie2024, Liu2024, Ban2024}.

From the above discussion, and given that the underlying quantum theory provides two types of solutions, a fundamental question arises: which solution is superior? {In this paper, we attempt to provide an answer from the perspective of mass inflation. The Strong Cosmic Censorship Conjecture (SCCC), which is closely related to mass inflation, states in Christodoulou’s modern formulation} \cite{Christodoulou1999} that even as a weak solution to the Einstein field equations, the metric generally cannot be extended beyond the Cauchy horizon. For asymptotically flat spacetime, the blueshift effect significantly amplifies linear perturbations outside the BH, ultimately transforming the Cauchy horizon into a mass inflation singularity. This is characterized by the unbounded growth of the quasilocal Hawking mass in a perturbed BH \cite{Matzner1979,Poisson1989,Ori1991,Carballo-Rubio2021,Iofa2022}, rendering the metric inextendible, which is consistent with the SCCC. However, certain exceptions to this scenario do exist (see \cite{Carballo-Rubio2022}). {If we also find a mass inflation singularity in the covariant quantum black hole, it would mean that this quantum black hole does not resolve the divergence problem inside the black hole and may therefore be flawed.} Moreover, most of the previous loop quantum BH solutions are not covariant, which is crucial for studying geodesics and perturbations \cite{Bolowald2015}. The solution, even with small perturbations, is usually no longer a solution of the same quantum-corrected Hamiltonian constraint. In contrast, the BH configurations obtained in \cite{Zhang2024, Bojowald2024,Zhang2024-12} preserve diffeomorphism invariance, providing a robust framework for addressing issues related to perturbations. {Therefore, we aim to investigate whether mass inflation exists in this newly proposed covariant quantum black hole.}

On the other hand, in asymptotically de Sitter (dS) spacetime, the stability of the Cauchy horizon is closely related to the decay rate of the dominant quasinormal modes (QNMs) \cite{Berti2009, Kokkotas1999, Destounis2024}. {Both scalar fields \cite{Cardoso2018, Cardoso2018PRD} and charged fermions \cite{Destounis2019} can lead to violation of the SCCC in RNdS spacetime. The same conclusion has also been observed for massless scalar perturbations in higher-dimensional RNdS spacetime \cite{Liu2019}. Charged accelerating BHs, sharing similar causal structures with RNdS spacetime, also exhibit SCCC violation \cite{Destounis2020}. Moreover, for scalar field perturbations non-minimally coupled to the Einstein tensor in RNdS spacetime (i.e., in Horndeski theory), SCCC is violated as well \cite{Destounis2019jhep}. The above conclusions can essentially be attributed to the perturbations} outside the event horizon exhibiting a more rapid exponential decay, which causes the redshift effect to suppress the blueshift effect at the Cauchy horizon. However, {asymptotically de Sitter spacetimes} do not necessarily guarantee the violation of the SCCC. In Kerr-dS BH backgrounds, it is interesting to note that no violations of the SCCC have been found in the cases of linear scalar and gravitational perturbations \cite{Dias2018}. Thus, a natural question is whether this quantum BH complies with the SCCC in dS spacetime. 

{We organize the paper as follows. In Section \ref{flat spacetime}, we discuss the phenomenon of mass inflation in the covariant quantum BH in asymptotically flat spacetime. In Section \ref{de Sitter spacetime}, we present the metric in the form with a cosmological constant and attempt to investigate the validity of the SCC. Section \ref{summary} provides a summary and discussion.}

\section{MASS INFLATION} \label{flat spacetime}

Poisson and Israel \cite{Poisson1990} employed an idealized null dust model to more comprehensively demonstrate that the Cauchy horizon of the RN-dS BH is unstable and that mass inflation occurs. Here, their analysis indicates that, in addition to the ingoing radiation (which undergoes infinite blueshift), one of the key ingredients for mass inflation is the presence of outgoing radiation inside the BH, which acts as a trigger. In the context of a realistic BH formed by gravitational collapse, the outgoing radiation originates from the surface of the collapsing star, while the ingoing radiation comes from radiation being scattered back into the BH due to the effects of spacetime curvature.  However, we do not know the full field equations corresponding to LQG. To investigate the stability of the Cauchy horizon, we can use the generalized Dray-'t Hooft-Redmond (DTR) relation \cite{Brown2011,Cao2024}, one of whose advantages is that it is independent of the field equations.

Before going into more detail, we introduce a spherical coordinate \((t, x, \theta, \phi)\), which adapts the symmetry of the spacetime. Then, the effective quantum-corrected metric possesses a Cauchy horizon and takes the form:
	\begin{align}\label{metric}
		\mathrm{d}s^{2}=& -f(x)\mathrm{d}t^{2}+\frac{\mathrm{d}x^{2}}{f(x)}+x^{2}\mathrm{d}\Omega^{2},
		\\
        f(x)=&1-\frac{2M}{x}+\frac{\zeta^{2}}{x^{2}}\left(1-\frac{2M}{x}\right)^{2},\label{metric function}
	\end{align}
where \(d\Omega^2 = d\theta^2+\sin^2\theta d\phi^2\), $M$ represents the mass  and $\zeta$ is a quantum parameter.
\begin{figure}[h!]
		\centering
		\includegraphics[scale=1.3]{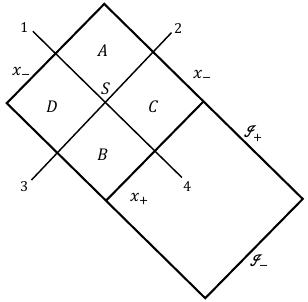}
		\caption{Penrose diagram, which provides a simplified depiction of the spacetime outside the BH and between the two horizons. The null shells collide at the two-dimensional sphere represented by the point $S$, dividing the spacetime inside the BH into four regions, where each one of them has its mass function.}
		\label{null shell}
\end{figure}

For simplicity, we will study a model that does not lose the essence of physics. In the spherically symmetric quantum spacetime described by eq.(\ref{metric}), as shown in Figure \ref{null shell}, the radiation inside the BH is replaced by radially propagating null shells. Lines 1 and 4 represent the ingoing null shell, while lines 2 and 3 are the outgoing null shell. These lines collide on the two-dimensional sphere $S$, dividing the spacetime between the horizons $x_{-}$ and $x_{+}$ into four regions: $A$, $B$, $C$, and $D$, each with its mass function, denoted as $M_A, M_B, M_C$ and $M_D$. They depend on {the advanced time \(v\equiv t+x_*\), where \( x_* \) is the tortoise coordinate defined by 
\begin{align}
	\frac{dx_*}{dx}=\frac{1}{f(x)}.
	\end{align} 
In other words, different \(v\) not only corresponds to a different position of the ingoing shell, but also to a different spacetime. However, each mass function remains unchanged throughout its corresponding region.} Region $B$ is connected to the asymptotically flat region outside the event horizon. Therefore, the mass of the BH observed by an external observer is \( M_B \). Due to the energy of the inner shell, the mass function undergoes a jump across the shell, which follows the relation given below \cite{Price1972}
  \begin{align}\label{mass relationship}
	{M_C\propto M_B+\frac{1}{v^p}}.
  \end{align}
Even though this conclusion is derived from classical BH theory, we expect it to be applicable in this context because the effects of quantum corrections should primarily play a role in the ultraviolet region. 

Based on the known generalized DTR relation, where the details are provided in \cite{Brown2011,Cao2024}, the four metric functions can be easily connected by 
	\begin{align}\label{DTR}
		|f_A|=\frac{|f_C|}{|f_B|}\cdot|f_D|.
	\end{align}
The subscript of $f$ indicates the corresponding spacetime provided previously in Figure \ref{null shell}. Here, we assume that the spacetime in each region is described by the corresponding metric function with its mass function. For example, the entire Region $A$ can be described by 
\begin{align}\label{f_A}
	f_A(x)=1-\frac{2M_A}{x}+\frac{\zeta^{2}}{x^{2}}\left(1-\frac{2M_A}{x}\right)^{2}.
\end{align}
It is worth noting that \(x_-\) and \(x_+\) are the roots of \(f_B(x)\), with the Cauchy horizon located at \(x = x_-\) and the event horizon at \(x = x_+\). If we rewrite $f_B(x)$ using the coordinates $x_-$ and $x_+$, we obtain
	\begin{align}
		f_B(x)=\frac{(2x_{-}-3x_{+})(x-x_{-})}{x_{-}^{2}}+O\left(x-x_{-}\right)^{2},\,\, x\rightarrow x_-, 
	\end{align}
which gives
	\begin{align}\label{behavior of f_B(x)}
		f_B(x)\propto x-x_- \mathrm{~,~~~~~}  x\rightarrow x_-.
	\end{align}
Since we are now focusing on the case where the ingoing shell is almost coincident with the Cauchy horizon, i.e., $v \rightarrow +\infty$ (see, e.g., \cite{Cao2024}), from eq. (\ref{mass relationship}) we have
    \begin{align}
    &f_C(x)=1-\frac{2M_C}{x}+\frac{\zeta^{2}}{x^{2}}\left(1-\frac{2M_C}{x}\right)^{2}\nonumber\\
    	&=f_B(x)-\frac{2\left(x^{3}+2x\zeta^{2}-4M_B\zeta^{2}\right)}{x^{4}}v^{-p}+O\left(v^{-p}\right)^2, \label{eq:fc}   	
    \end{align}
where the second term is {nonzero} as $x\rightarrow x_-$. Using the asymptotic behaviors
	\(x-x_-\propto e^{-\kappa_-v}\) {(see, e.g., \cite{Cao2024})}, where \(\kappa_-\) is the surface gravity of the Cauchy horizon,
 and eq.(\ref{behavior of f_B(x)}), we obtain that
	\begin{align}\label{fc/fb}
		\frac{f_C}{f_B}\sim f_B^{-1}v^{-p}\sim v^{-p}e^{\kappa_-v}\to+\infty \mathrm{~,~~~~~} v \rightarrow +\infty.  
	\end{align}
Generally, since $M_A$, $M_D$, and $M_B$ are all different from each other, $x = x_-$ is only a root of $f_B(x)$ rather than $ f_A(x)$ or $f_D(x)$, which means $f_D(x)$ is {nonzero. Furthermore, it can be seen from eqs. (\ref{DTR}) and (\ref{fc/fb}) that $f_A(x)$ must be infinite. Finally, according to} eq.(\ref{f_A}), the only way for $f_A(x)$ to diverge is that:
	\begin{align}
		M_A\rightarrow +\infty.
	\end{align}
In other words, mass inflation occurs near the Cauchy horizon, which prevents the extension of the spacetime beyond the Cauchy horizon, preserving the SCCC. Note that \(M_A\) is a parameter of the spacetime that is only accessible deep inside the BH and is not observable from the outside \cite{Brown2011}.
A situation similar to that can be seen in the quantum Oppenheimer-Snyder model \cite{Cao2024}.

\section{VALIDITY OF THE STRONG COSMIC CENSORSHIP CONJECTURE}\label{qnm and SCC}
\label{de Sitter spacetime}

The premise for studying the SCCC in dS spacetime is extending the effective metric (\ref{metric}) via the inclusion of a cosmological constant. As discussed in \cite{Zhang2024}, {classical general relativity is modified within the Hamiltonian formulation into an effective semiclassical gravity model while preserving 4-dimensional diffeomorphism covariance.} The spherical symmetry helps us reduce the {general relativity} to dilaton gravity on the 2-manifold \(\mathcal{M}_2=\mathbb{R}\times\Sigma\) {with \(\Sigma\) being the spatial manifold. The canonical variables on \(\Sigma\) satisfy \(\{K_1(x),E^1(y)\}=2\delta(x,y)\mathrm{~}\) and \(\mathrm{~} \{K_2(x),E^2(y)\}=\delta(x,y)\).} The corresponding 4-dimensional metric on the 4-manifold $\mathcal{M}_2\times\mathbb{S}^2$ {(with \(\mathbb{S}^2\) denoting the 2-sphere)} reads
	\begin{align}
		\mathrm{d}s^2=-N^2\mathrm{d}t^2+\frac{(E^2)^2}{\mu E^1}(\mathrm{d}x+N^x\mathrm{d}t)^2+E^1\mathrm{d}\Omega^2,	
	\end{align}
where \(N\) is a lapse function, \(N^x\) a shift vector, and \(\mu\) is the factor caused by the quantum gravity effects. {The dynamics of the canonical variables is encoded in the diffeomorphism constraint \(H_x\) and the Hamiltonian constraint \(H_{\mathrm{eff}}\).} The diffeomorphism constraint retains its classical form
	\begin{align}
		H_x=-K_1\partial_xE^1/2+E^2\partial_xK_2,
	\end{align} 
	and, {due to quantum corrections,} the Hamiltonian constraint takes the form
	\begin{align}\label{Hamiltonian constraint}
			&H_{\mathrm{eff}}=-\frac{E^2}{2\sqrt{E^1}}-\frac{K_1E^1}{2\zeta}\sin\left(\frac{2\zeta K_2}{\sqrt{E^1}}\right)\nonumber \\
			&-\frac{3\sqrt{E^1}E^2}{2\zeta^2}\sin^2\left(\frac{\zeta K_2}{\sqrt{E^1}}\right)+\frac{K_2E^2}{2\zeta}\sin\left(\frac{2\zeta K_2}{\sqrt{E^1}}\right)\nonumber \\
			&+\frac{\left(\partial_xE^1\right)^2}{8\sqrt{E^1}E^2}e^{\frac{2\mathrm{i}\zeta K_2}{\sqrt{E^1}}}+\frac{\sqrt{E^1}}2\partial_x\left(\frac{\partial_xE^1}{E^2}\right)e^{\frac{2\mathrm{i}\zeta K_2}{\sqrt{E^1}}} \nonumber\\
			&+\frac{i\zeta E^2}4\left(\frac{\partial_xE^1}{E^2}\right)^2\left(\frac{K_1}{E^2}-\frac{K_2}{E^1}\right)e^{\frac{2\mathrm{i}\zeta K_2}{\sqrt{E^1}}}\nonumber\\
            &+\frac{\sqrt{E^1}E^2 \Lambda}{2},
	\end{align}
	{where \(H_{\mathrm{eff}}\) reduces to the classical Hamiltonian constraint when the quantum parameter \(\zeta\) is set to zero.} 
	The effective theory has a Dirac observable, which is essentially the BH mass given by
	\begin{align}\label{Meff}
		M_{\mathrm{eff}}&=\frac12\sqrt{E^1}\left(1-\frac{\Lambda E^1}{3}\right)+\frac{(E^1)^{3/2}}{2\xi^2}\sin^2\left(\frac{\zeta K_2}{\sqrt{E^1}}\right)\nonumber\\
        &-\frac{\sqrt{E^1}}{8}\left(\frac{\partial_x E^1}{E^2}\right)^2\mathrm{e}^{\frac{2\mathrm{i}\zeta K_2}{\sqrt{E^1}}}.
	\end{align}
    Here, eqs. (\ref{Hamiltonian constraint}) and (\ref{Meff}) correspond to one of the solutions with a cosmological constant from {eqs. (6) and (7)} of Ref. \cite{Zhang2024}, where \(\mu\) can be determined to be \(1\). {Following the steps in \cite{Zhang2024} (or referring to \cite{Zhang2024-12} for more details), the areal gauge \(E^1(x) = x^2\) is chosen to fix the gauge generated by the diffeomorphism constraint. Then the lapse function \(N\) and shift vector \(N^x\) are determined by solving 
    \begin{align}
    	\{E^I(x), H^{(1)}_{\mathrm{eff}}[N] + H_x[N^x]\} = 0.
    \end{align}  
    Further imposing \(N^x = 0\) and \(H_{\mathrm{eff}} = 0\), \(E^2(x)\) is uniquely determines.} Finally, the effective metric with a cosmological constant is given by
    \begin{align}\label{effective metric}
		\mathrm{d}s_\Lambda^{2}=-f_\Lambda(x)\mathrm{d}t^{2}+\frac{\mathrm{d}x^{2}}{f_\Lambda(x)}+x^{2}\mathrm{d}\Omega^{2},
	\end{align}
     where
	{\begin{align}\label{f_Lambda}
			f_\Lambda(x)=&\left(1-\frac{2M}x\right)\left(1-\frac{2\zeta^2\Lambda}3\right) +\frac{\zeta^2}{x^2}\biggl(1-\frac{2M}x\biggr)^2\nonumber \\ 
			&-\left(1-\frac{\zeta^2\Lambda}3\right)\frac{x^2\Lambda}3.
	\end{align}}
    Here, to maintain consistency with the notation used in other sections of this work, we have denoted the BH's effective mass \( M_{\text{eff}} \) as \( M \). 

    Let us analyze the relation between BH parameters and the positions of the horizons, and determine the parameter intervals. When the spacetime contains inner and outer horizons along with a cosmological horizon \(x_c\), their connections to the quantum BH parameters are described by
    \begin{align}
	\Lambda =&\frac{3}{{x_{c}}^{2}+{x_{c}x_{+}}+{x_{+}}^{2}}, \\
	 M =&\frac{(x_{c}+x_{+})x_{c}x_{+}}{2({x_{c}}^{2}+x_{c}x_{+}+{x_{+}}^{2})}, \\
	\zeta^2=&\frac{(x_c^2+x_cx_-+x_+^2)x_-^3}{(x_c-x_-)(x_+-x_-)(x_c+x_++x_-)}.
	\end{align}
    Through simple analysis, it can be understood that \( M \) and \( \Lambda \) are determined by \( x_+ \) and \( x_c \). When they are fixed, as \( x_- \) approaches \( x_+ \), \( \zeta \) increases monotonically. Finally, as the BH approaches the near-extremal limit, \( \zeta \) tends to infinity.  However, we argue that \( \zeta \) admits an upper bound, and the justification for this will be discussed below. Within the framework of covariant LQG, \( \zeta \) is defined as
    {
   	\begin{align}
    			\zeta=\sqrt{4\sqrt{3}\pi\gamma\ell_p^2},
    \end{align}} 
    with \(\gamma\) being the Barbero-lmmirzi parameter and \(\ell_p\) the Planck length. To recover the Bekenstein–Hawking area law for quantum BHs, the Immirzi parameter is typically chosen as \(\gamma \approx 0.237\). Similarly, other methods suggest a different value \(\gamma \approx 0.274\) (see Refs. \cite{Bambi2023, Vyas2022}). It is worth noting that even in the context of matching BH thermodynamics, its exact value remains theoretically uncertain. This inherent ambiguity in \(\gamma\) naturally extends to the related quantum parameter \(\zeta\), leading to a corresponding uncertainty in their allowable ranges. In this work, we set \(\gamma \approx 0.274\), which determines \textcolor{black}{\(\zeta \approx 2.4420\)} in Planck units.  Furthermore, we introduce the minimum BH mass \(M_{min}\). In the effective LQG, the smallest nonzero eigenvalue of the area operator is \(\Delta=4 \sqrt{3} \pi \gamma \ell_p^2\). By setting \(\Delta\) equal to the horizon area of the BH,  we have \textcolor{black}{\(M_{min} \approx 0.3445\)} in Planck units. For convenience, we normalize the mass, allowing all physical quantities to be expressed in terms of \(M\) in a dimensionless form, such as \(\zeta \to \zeta / M\) and \(\Lambda \to \Lambda M^2\). With the presence of \(M_{min}\), the dimensionless quantity satisfies \textcolor{black}{\(0<\zeta/M \lesssim 7.090 \)}. Moreover, to guarantee the existence of at least three horizons, the condition \textcolor{black}{\(0 <\Lambda M^2 \lesssim 0.111\)} must be satisfied.

In the effective quantum Oppenheimer-Snyder model, it has been proven that a near-extremal quantum BH in dS spacetime violates the SCCC. This is due to the remnant perturbation fields outside the BH do not decay polynomially but rather exponentially \cite{Shao2024}. Motivated by these results, we aim to investigate whether the SCCC is also violated in the covariant quantum model. Assuming that the effective action of LQG with a classical scalar field can be written as \cite{Shao2024}  
\begin{align}
	S=\int d^4x\sqrt{-g}\left(\frac{\mathcal{R}-2\Lambda}{16\pi}+L_\hbar(g)-\frac{1}{2}\nabla_\mu\Psi\nabla^\mu\Psi\right),
\end{align}	
	where \(\mathcal{R}\) is the Ricci scalar, \(\Psi\) is a massless scalar field, and \(L_{\hbar} \) represents the quantum correction terms related to spacetime. When we vary the action with respect to the metric \(g\), we should obtain the field equations modified by quantum corrections \textcolor{black}{
	\begin{align}\label{field equations}
			G^\hbar_{\mu\nu}(g) = 8\pi T_{\mu\nu}^{(sc)},
	\end{align}
	where \(G^\hbar_{\mu\nu}(g)\) is the quantum-corrected Einstein tensor and \(T_{\mu\nu}^{(sc)}\) is the energy-moment associated with the scalar field:
	\begin{align}
			T_{\mu\nu}^{(sc)} = \partial_\mu \Psi \partial_\nu \Psi - \frac{1}{2} g_{\mu\nu} \partial_\alpha \Psi \partial^\alpha \Psi.
	\end{align}}  

    \begin{figure*}[htbp]
	\suppressfloats[t]      
	\centering  
	\begin{subfigure}[b]{0.47\textwidth}   
		\includegraphics[width=\textwidth]{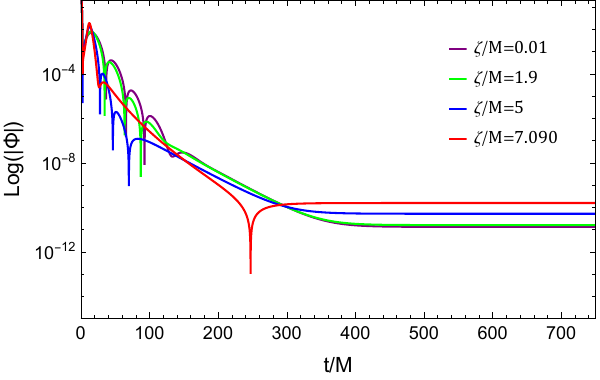}  
		\caption{ \textcolor{black}{Fixed $l=0$, $\Lambda M^2=0.001$, varying $\zeta/M$}}  
		\label{temporal evolution a}  
	\end{subfigure}  
	\begin{subfigure}[b]{0.47\textwidth}  
		\includegraphics[width=\textwidth]{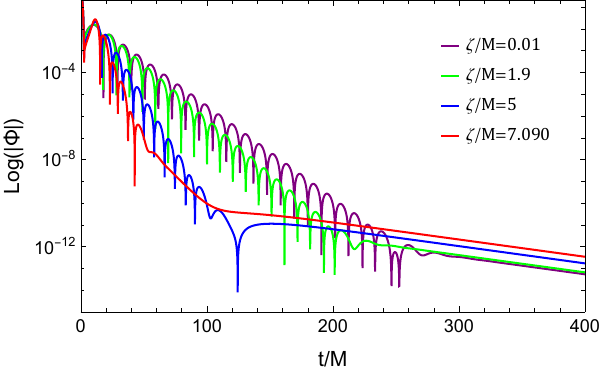}  
		\caption{ \textcolor{black}{Fixed $l=1$, $\Lambda M^2=0.001$, varying $\zeta/M$}}  
		\label{temporal evolution b}  
	\end{subfigure}  
	\begin{subfigure}[b]{0.47\textwidth}    
		\includegraphics[width=\textwidth]{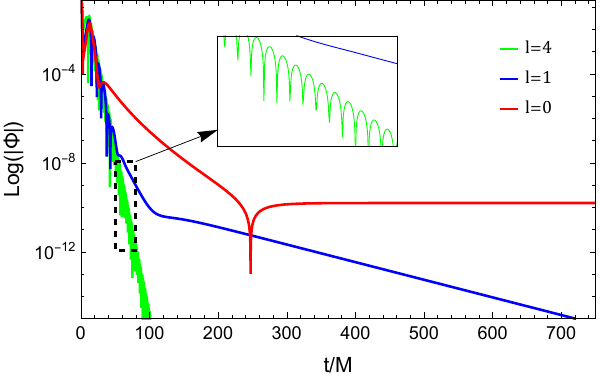}  
		\caption{ \textcolor{black}{Fixed $\zeta/M=7.090$, $\Lambda M^2=0.001$, varying $l$}}  
		\label{temporal evolution c}  
	\end{subfigure}  
	\begin{subfigure}[b]{0.47\textwidth}   
		\includegraphics[width=\textwidth]{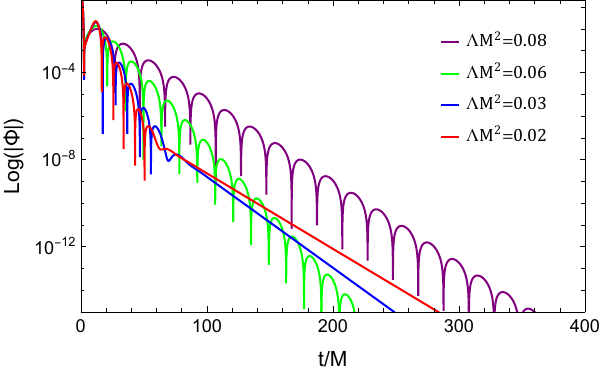}  
		\caption{ \textcolor{black}{Fixed $l=1$, $\zeta/M=7.090$, varying $\Lambda M^2$}}  
		\label{temporal evolution d}  
	\end{subfigure}  
	
	\caption{\textcolor{black}{Temporal evolution of massless scalar perturbations. Subfigure legends indicate the varied parameters.}}\label{temporal evolution}    
\end{figure*}

According to the formulation by Christodoulou \cite{Christodoulou2009}, the SCCC states that, for general smooth initial data, the metric cannot be extended across the Cauchy horizon as a weak solution. \textcolor{black}{If the solution of Eq.(\ref{field equations}) is extendible beyond the Cauchy horizon, the condition for the existence of weak solutions must be satisfied:
\begin{align}\label{weak solutions of field equations}
		\int_{\mathcal{V}} d^4x \sqrt{-g} [G^\hbar_{\mu\nu}(g) - 8\pi T_{\mu\nu}^{(sc)}] \xi = 0,
\end{align}
where \(\xi\) is a smooth and compactly supported test function, and \(\mathcal{V}\) is a small neighborhood near the Cauchy horizon. 
The existence of weak solutions requires the result of the integral to be bounded. For the classical Einstein tensor \(G_{\mu\nu}(g)\), this requires the Christoffel symbols to belong to a locally square-integrable function space. However, for \(G^\hbar_{\mu\nu}(g)\), since its exact expression is not yet well understood in the context of effective quantum gravity, we assume it satisfies local integrability. Without this assumption, we would not be able to proceed with the subsequent discussion. As stated in \cite{Shao2024}, if the \(T_{\mu\nu}^{(sc)}\) consisting of the square of its first derivative for the scalar field can be integrable at the Cauchy horizon, then QNMs must satisfy}  
\textcolor{black}{\begin{align}
	\beta\equiv\frac{\mathrm{inf}\{-\mathrm{Im}(\omega)\}}{\kappa_-}>\frac{1}{2},
\end{align}
	where \(\mathrm{inf}\) means infimum and \(\mathrm{Im}(\omega)\) refers to the imaginary parts of the QNMs of the perturbations in the external region of the BH.} In other words, if there exists a lowest lying QNM such that \( \beta < 1/2 \), the SCCC holds. Now consider \(\Psi\) in the aforementioned background (\ref{effective metric}), the equation of motion is given by
	\begin{align}\label{equation of motion}
		\frac{1}{\sqrt{-g}}\partial_{\mu}\left(\sqrt{-g}g^{\mu\nu}\partial_{\nu}\Psi\right)=0.
	\end{align}
Through the separation of variables by spherical harmonic functions \(\Psi=\Phi(t,x)Y_l(\theta,\phi)/x\), eq.(\ref{equation of motion}) can be reduced to 
\begin{align}\label{continuous equation}
	-\frac{\partial^2\Phi}{\partial t^2}+\frac{\partial^2\Phi}{\partial x_*^2}-V(x)\Phi=0.
	\end{align}
Here, the effective potential is
\begin{align}
	V(x)=f_\Lambda(x)\left[\frac{l(l+1)}{x^2}+\frac{f_\Lambda'(x)}x\right],
\end{align}
with \(l\) the angular quantum number. Further, assuming that \(\Phi=e^{-i\omega t}\psi(x)\), the resulting radial equation takes the Schr\"{o}dinger wavelike form
	\begin{align}\label{wave-like equation}
		\frac{\partial^2\psi}{\partial x_*^2}+\left(\omega^2-V(x)\right)\psi=0.
	\end{align}

To investigate whether the late-time behavior of the scalar field also exhibits exponential decay, we employ the finite difference method, replacing the continuous equation (\ref{continuous equation}) with discrete differences and performing integration. Consequently, the temporal evolution can be obtained. Calculating QNMs is essentially obtaining the intrinsic frequencies of eq.(\ref{wave-like equation}). In dS spacetime, the boundary conditions near the event horizon and the cosmological horizon require that waves propagate purely toward each respective horizon. In our work, we primarily use the pseudospectral method \cite{Jansen2017,Mamani2022}, supplemented by the direct integration method \cite{PANI2013} to verify QNMs.

\begin{table*}[htb]  
	\setlength{\tabcolsep}{7pt}  
	\centering  
	\caption{\textcolor{black}{The fundamental QNMs for different \(\zeta/M\) and \(\Lambda M^2\) at \(l=0,1\). The upper and lower rows data are obtained by the pseudospectral method and the direct integration method, respectively.}}  
	\renewcommand{\arraystretch}{1.5}    
	\textcolor{black}{  
		\begin{tabular}{M{1.5em} c cccc}  
			\toprule  
			\(l\) & \(\Lambda M^2\) & \(\zeta/M = 1.9\) & \(\zeta/M = 5\) & \(\zeta/M = 7.090\) \\   
			\midrule    
			\multirow{6}{*}{\vspace{-19.5pt} 0}  
			& \multirow{2}{*}{0.01}   
			& \(-0.1156475 i\) & \(-0.0950471i\)& \(-0.0826510i\)\\ \cmidrule(lr){3-5}  
			& & \(-0.1156468 i\) & \(-0.0950360i\)& \(-0.0826510i\)\\   
			\cmidrule(lr){2-5}  
			& \multirow{2}{*}{0.05}   
			& \(0.0726016 - 0.1056059 i\) & \(-0.1422525i\) & \(-0.1118924i\)\\ \cmidrule(lr){3-5}  
			& & \(0.0726016 - 0.1056059 i\) & \(-0.1422525i\) & \(-0.1118924i\)\\   
			\cmidrule(lr){2-5}  
			& \multirow{2}{*}{0.10}   
			& \(0.0172417 - 0.0553389 i\) & \(0.0158140 - 0.0571378 i\) & \(0.0139424 - 0.0588729i\)\\  \cmidrule(lr){3-5} 
			& & \(0.0172417 - 0.0553389 i\) & \(0.0158140 - 0.0571378 i\) & \(0.0139424 - 0.0588729i\)\\   
			\midrule  
			\multirow{6}{*}{\vspace{-19.5pt} 1}  
			& \multirow{2}{*}{0.01}   
			& \(-0.0576674 i\) & \(-0.0574548i\)& \(-0.0572142i\)\\  \cmidrule(lr){3-5}  
			& & \(-0.0576674 i\) & \(-0.0574548i\)& \(-0.0572142i\)\\   
			\cmidrule(lr){2-5}  
			& \multirow{2}{*}{0.05}   
			& \(0.2105812 - 0.0835571 i\) & \(-0.1236650i\)& \(-0.1187440i\)\\  \cmidrule(lr){3-5} 
			& & \(0.2105812 - 0.0835571 i\) & \(-0.1236650i\)& \(-0.1187440i\)\\   
			\cmidrule(lr){2-5}  
			& \multirow{2}{*}{0.10}   
			& \(0.0818801 - 0.0317033 i\) & \(0.0834996 - 0.0344703i\)& \(0.0852071 - 0.0376948i\)\\  \cmidrule(lr){3-5} 
			& & \(0.0818801 - 0.0317033 i\) & \(0.0834996 - 0.0344703i\)& \(0.0852071 - 0.0376948i\)\\   
			\bottomrule  
		\end{tabular}  
	}   
	\label{tab:quasinormal_modes}  
\end{table*}
\begin{table}[htb]  
	\setlength{\tabcolsep}{7pt}
	\centering  
	\caption{\textcolor{black}{The \(-\mathrm{Im}(\omega)/\kappa_-\) values obtained by the pseudospectral method are presented for different \(\zeta/M\) and \(\Lambda M^2\) at \( l=0, 1\).}}  
	\textcolor{black}{
		\begin{tabular}{cccccc}  
			\toprule  
			\(l\) &\(\Lambda M^2\) & \(\zeta/M=1.9\) & \(\zeta/M=5\) & \(\zeta/M=7.090\) \\ \midrule    
			\multirow{3}{*}{\vspace{-11pt} 0}  
			& \(0.01\)   & \(0.1252543\) & \(0.2515794\) & \(0.2651810\)\\  \cmidrule(lr){2-5}
			& \(0.05\)   & \(0.0616234\) & \(0.4248653\) & \(0.4259152\)\\   \cmidrule(lr){2-5}
			& \(0.10\)   & \(0.0639121\) & \(0.2071366\) & \(0.3069090\)\\\midrule
			\multirow{3}{*}{\vspace{-11pt} 1}  
			& \(0.01\)   & \(0.0624578\) & \(0.1520766\)& \(0.1835683\)\\  \cmidrule(lr){2-5}
			& \(0.05\)   & \(0.0930460\) & \(0.3693502\)& \(0.4519953\)\\   \cmidrule(lr){2-5} 
			& \(0.10\)   & \(0.0366149\) & \(0.1249619\)& \(0.1965058\)\\
			\bottomrule  
	\end{tabular}}  
	\label{tab:ratio}  
\end{table}

	\section{NUMERICAL RESULTS}\label{numerical results}

In this section, we present the results of our numerical calculation. \textcolor{black}{Let us first discuss the effects of angular quantum number \(l\), the cosmological constant \(\Lambda M^2\), and the quantum parameter \(\zeta/M\) on the late-time behavior of the temporal evolution.} As a demonstration, Figure \ref{temporal evolution} shows the temporal evolution of massless scalar fields simulated by a Gaussian wave packet. \textcolor{black}{It can be seen from subfigures \ref{temporal evolution a} and \ref{temporal evolution b} that, under different values of \(\zeta/M\), the late-time behavior of the solutions for both \(l=0\) and \(l=1\) in eq.(\ref{continuous equation}) appears as straight lines in the logarithmic plot. This indicates that the late-time behavior in asymptotically dS spacetime indeed overpower the quasinormal frequency and exhibit exponential decay, eventually resulting in purely imaginary quasinormal modes. Additionally, when quantum parameters are different, the late-time tails are almost parallel, suggesting that the quantum parameter has a negligible effect on the late-time behavior of the scalar field evolution.} \textcolor{black}{On the other hand, as shown in the subfigure \ref{temporal evolution c}, for a propagating scalar field with larger \(l\), a tail is replaced by the longest-lived quasinormal frequencies, which persist indefinitely. The subfigure \ref{temporal evolution d} further demonstrates that for \(l=1\), no tail forms when \(\Lambda M^2\) is large enough. Overall, at late times, all dominant modes decay either through a tail or through long-lived quasinormal ringing.}

\textcolor{black}{Typically, we can assess the validity of the SCCC in the following way. First, by solving the QNM problem in specific regions of the parameter space (i.e., near the “boundaries” or “corners”). In these regions, analytic expressions for the QNM frequencies can often be obtained through approximation methods, allowing us to identify each family of modes. Then, by continuously varying the parameters, we can track these families and thus extend their classification throughout the entire parameter space, comparing and selecting the least damped QNM. This approach is very effective for identifying violations of the SCCC, but to establish the validity of the SCCC it suffices to find a single QNM such that \(-\mathrm{Im}(\omega)/\kappa_- < 1/2\). Fortunately, for the case we are about to discuss, it is not necessary to strictly classify all QNMs or to always find the least-damped QNM. In fact, it is sufficient to select the QNMs with the smaller damping among \(l=0\) and \(l=1\) modes, and it can be found that \(-\mathrm{Im}(\omega)/\kappa_-<1/2\) holds throughout the entire parameter range, which implies the validity of the SCCC.}
 
To demonstrate the validity of SCCC, we focus only on the fundamental mode with \( n=0 \), which corresponds to the longest-lived mode \textcolor{black}{without regard to \(l\),} i.e., to the one with smallest imaginary part (in absolute value) of the frequency. \textcolor{black}{In the table (\ref{tab:quasinormal_modes}) below, we present some of the calculated QNMs. The values calculated using the two numerical methods are almost identical. It is obvious that the QNMs for \(\Lambda M^2=0.01\) are purely imaginary whereas for larger \(\Lambda M^2\) oscillatory QNMs are obtained. However, for larger values of \(\zeta/M\), the QNMs tend to change from a complex mode to a purely imaginary one.}

\begin{figure}[htbp]      
	\centering  
	\includegraphics[scale=0.8]{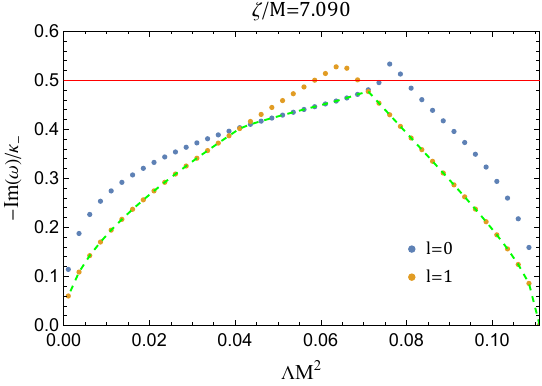} 
	\caption{\textcolor{black}{The \(-\mathrm{Im}(\omega)/\kappa_-\) for \(l =0,1\) with \(\zeta/M=7.090\).}}  
	\label{fig:ratio}  
\end{figure}
\begin{figure}[htbp]      
	\centering  
	\includegraphics[scale=0.75]{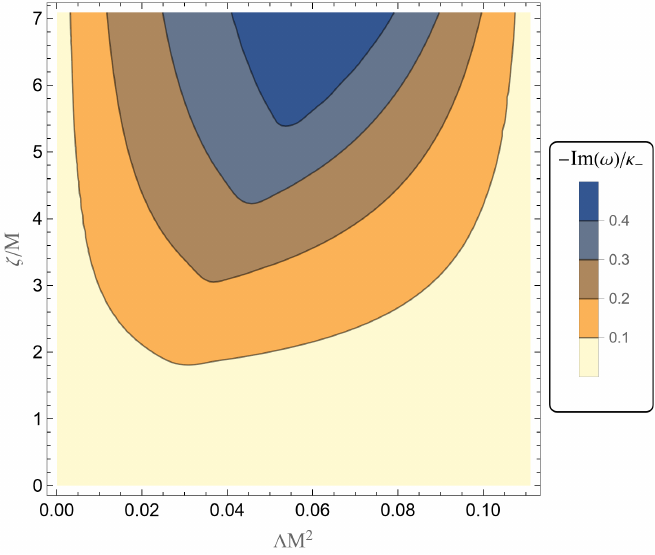} 
	\caption{ The density plot of \textcolor{black}{\(-\mathrm{Im}(\omega)/\kappa_-\)} within the effective range of \(\zeta/M\) and \(\Lambda M^2\).}  
	\label{DensityPlot}  
\end{figure}
\textcolor{black}{Since the reliability of the pseudospectral method has already been verified against the direct integration method in Table \ref{tab:quasinormal_modes}, all subsequent calculations of \(-\mathrm{Im}(\omega)/\kappa_-\) are performed using the pseudospectral method. In Table \ref{tab:ratio}, it can be seen that when \(\Lambda M^2 = 0.05\) and \(\zeta/M\) reaches its upper limit, \(-\mathrm{Im}(\omega)/\kappa_-\) is quite close to 1/2, indicating that careful computation is required for nearby parameter regions. In Figure \ref{fig:ratio}, the \(-\mathrm{Im}(\omega)/\kappa_-\) is exhibited for both \(l=0\) and \(l=1\) across the entire range of \(\Lambda M^2\) with \(\zeta/M\) fixed, and the green dashed line represents the smaller values. It can be seen that within this parameter range, the SCCC always holds. However, if one considers only \(l=0\) or \(l=1\) individually, there are cases where \(-\mathrm{Im}(\omega)/\kappa_->1/2\). We extend this approach to the entire parameter space and obtain a density plot of \(-\mathrm{Im}(\omega)/\kappa_-\), it is evident that all \(-\mathrm{Im}(\omega)/\kappa_-\) values are smaller than \(1/2\)}. \textcolor{black}{It should be noted that while we cannot guarantee the \(l=0,1\) modes always dominate the late-time ringdown, if another mode has a smaller absolute value of the imaginary part, the \(-\mathrm{Im}(\omega)/\kappa_-\) would be even smaller. Therefore, the above ratio would certainly be smaller than \(1/2\) and we can conclude that the SCCC is upheld.}

\section{Summary and conclusions} \label{summary}

In the present work, we discuss the stability of the Cauchy horizon and the SCCC in spherically symmetric quantum BHs that satisfy general covariance. The results indicate that mass inflation occurs in asymptotically flat spacetime, suggesting that the Cauchy horizon is unstable. Furthermore, we extended the effective metric to include a cosmological constant via eq. (\ref{effective metric}). Numerical calculations reveal that the SCCC holds throughout the entire parameter space.
	
The original intention of quantum gravity is to avoid singularities in spacetime. Although the quantum BH is regular at \( x=0 \), we have confirmed that this quantum BH model also exhibits mass inflation at the Cauchy horizon.  Additionally, this phenomenon has not been observed for the first time in quantum BHs \cite{Carballo-Rubio2021,Brown2011,Cao2024}. As mass inflation drives the inner horizon to evolve into a null singularity, this divergent behavior essentially eliminates the possibility of extending the metric beyond the Cauchy horizon in the form of a weak solution. Thus, mass inflation plays a crucial role in maintaining the validity of the SCCC. In other words, while mass inflation goes against the original intention of constructing quantum BHs, it allows the SCCC to hold. A more reasonable model should both ensure the absence of divergent behavior in spacetime and the validity of the SCCC. Although we only explored one metric (\ref{metric}) in \cite{Zhang2024}, a similar analysis can also be conducted for the other metric as well. Currently, our results suggest that another BHs solution without the Cauchy horizon presented in \cite{Zhang2024-12} is a more reasonable alternative. 

Regarding the discussion of SCCC in dS, it must be emphasized that a more reasonable approach would be to analyze the quantum scalar field within the framework of LQG. The current work provides only a preliminary exploration and is limited to a classical scalar field probe, which does not fully account for the quantum nature of BHs. In contrast, within a semiclassical framework, Ref. \cite{Hollands2020} considers a quantum scalar field in RN-dS spacetime, and shows that quantum effects can restore the validity of the SCCC. Further discussions on quantum scalar fields at the Cauchy horizon can be found in Refs. \cite{Hintz2024,Arrechea2024}. We look forward to future studies of coupled scalar fields in the context of quantum spacetimes.

\section*{Acknowledgements}
This work is supported by National Natural Science Foundation of China (NSFC) with Grant No. 12275087. M.B. is supported by proyecto interno UCM-IN-25202.

\end{document}